\documentclass{elsart}
\usepackage{graphicx}
\usepackage{amsmath,amssymb}
\makeatletter
  \def\mathcomposite{%
     \@ifstar
        {\def\@mathcomposite@option{%
            \baselineskip\z@skip\lineskiplimit-\maxdimen}%
         \@mathcomposite}%
        {\let\@mathcomposite@option\offinterlineskip
         \@mathcomposite}}
  \def\@mathcomposite{%
     \@ifnextchar[\@@mathcomposite{\@@mathcomposite[0]}}
  \def\@@mathcomposite[#1]#2#3#4{%
     #2{\mathchoice
        {\@mathcomposite@{#1}{#3}{#4}\displaystyle{1}}%
        {\@mathcomposite@{#1}{#3}{#4}\textstyle{1}}%
        {\@mathcomposite@{#1}{#3}{#4}%
         \scriptstyle\defaultscriptratio}%
        {\@mathcomposite@{#1}{#3}{#4}%
         \scriptscriptstyle\defaultscriptscriptratio}}}
\def\@mathcomposite@#1#2#3#4#5{%
     \vcenter{\m@th\@mathcomposite@option
        \dimen@\f@size\p@\dimen@#1\dimen@\dimen@#5\dimen@
        \divide\dimen@ 18
        \edef\@mathcomposite@skipamount{\the\dimen@}%
        \ialign{\hfil$#4##$\hfil\cr
           #2\crcr
           \noalign{\vskip\@mathcomposite@skipamount}%
           #3\crcr}}}
\makeatother

\begin{document}
\begin{frontmatter}

\title{Indication of a strange tribaryon S$\mathbf{^+}$ from 
the $^4$He(stopped K$^-$,n) reaction
% \\ : as of 04-11-16
}

\author[label5,label3]{M.~Iwasaki},
\ead{masa@riken.go.jp}
\author[label4]{T.~Suzuki},
\author[label2]{H.~Bhang},
\author[label1]{G.~Franklin},
\author[label4]{K.~Gomikawa},
\author[label4]{R.S.~Hayano},
\author[label3]{T.~Hayashi\thanksref{cor1}},
\author[label3]{K.~Ishikawa},
\author[label6]{S.~Ishimoto},
\author[label5]{K.~Itahashi},
\author[label3]{T.~Katayama},
\author[label3]{Y.~Kondo},
\author[label5]{Y.~Matsuda},
\author[label3]{T.~Nakamura},
\author[label3]{S.~Okada\thanksref{cor2}},
\author[label6]{H.~Outa\thanksref{cor2}},
\author[label1]{B.~Quinn},
\author[label3]{M.~Sato},
\author[label4]{M.~Shindo},
\author[label2]{H.~So},
\author[label3]{T.~Sugimoto},
\author[label5]{P.~Strasser\thanksref{a3}},
\author[label4]{K.~Suzuki\thanksref{cor3}},
\author[label6]{S.~Suzuki},
\author[label3]{D.~Tomono\thanksref{a3}},
\author[label3]{A.M.~Vinodkumar},
\author[label4]{E.~Widmann\thanksref{cor4}},
\author[label5]{T.~Yamazaki},
\author[label3]{T.~Yoneyama}

\address[label5]{DRI, RIKEN, 
Wako-shi, Saitama, 351-0198, Japan}
\address[label3]{Department of Physics, Tokyo Institute of Technology, 
Ookayama, Meguro-ku, Tokyo 152-8551, Japan}
\address[label4]{Department of Physics, University of Tokyo, 
Hongo, Bunkyo-ku, Tokyo 113-0033, Japan}
\address[label2]{Department of Physics, Seoul National University, 
Shikkim-dong, Kwanak-gu, Seoul 151-742, Korea}
\address[label1]{Department of Physics, Carnegie Mellon University, 
Pittsburgh, PA 15213, USA}
\address[label6]{IPNS, KEK (High Energy Accelerator Research Organization), 
Oho, Tsukuba-shi, Ibaraki 305-0801, Japan\\}%
\thanks[cor1]{Present address: Department of Legal Medicine, Osaka University}
\thanks[cor2]{Present address: DRI, RIKEN}
\thanks[a3]{Present address: IMSS, KEK}
\thanks[cor3]{Present address: Physik-Department E12, 
Technische Universit\"at M\"unchen}
\thanks[cor4]{Present address: Stefan Meyer Institute for Subatomic Physics, 
Austrian Academy of Sciences, Boltzmanngasse 3, A-1090 Wien, Austria}

\begin{abstract}
We measured the neutron time-of-flight spectrum in the
$^4$He(${\rm stopped}K^{-}, n$) reaction by stopping 
negative kaons in a superfluid helium target.
   A clear enhancement was observed in the neutron spectrum, 
which indicates the formation of a strange tribaryon of charge +1 
with a mass and width of  
$M_{\rm{S}^+} = $ 3141 $\pm$ 3 (stat.) $^{+4}_{-1}$ (sys.) MeV/$c^2$ 
and $\mathnormal{\Gamma}_{{\rm{S}}^+} \lesssim 23 $ MeV/$c^2$. 
This state, denoted as S$^+$(3140), is about 25 MeV/$c^2$ higher 
than the previously observed S$^0$(3115) $T=1$.
\end{abstract}

\end{frontmatter}

% main text

\section{Introduction}

Very recently, we have discovered a strange tribaryon, S$^0$(3115), via the 
\begin{equation}
\label{eq:K-p-reaction}
    (K^{-}  {~^4}{\rm He})_{\rm atomic} \rightarrow {\rm{S}}^{0}(3115) + p
\end{equation}
reaction \cite{Suz04}.
   The observed state has isospin $T=1$, charge $Z=0$
and mass $M_{{\rm{S}}^{0}}$ $\sim$ 3117 MeV/$c^2$.

   Where is the isospin $T$ = 0 state? 
   It is also important to search for an isospin partner of the observed 
state, S$^0$(3115) $T$ = 1. 
   The S$^0$(3115) discovery was originally motivated 
by the theoretical prediction of a
deeply-bound kaonic state by Akaishi and Yamazaki 
at $M$ = 3194 MeV/$c^2$ with $T=0$ and $Z=1$ \cite{PRC02}, 
which can only be studied via the 
\begin{equation}
\label{eq:K-n-reaction}
	(K^{-} {~^4}{\rm He})_{\rm atomic} \rightarrow {\rm{S}}^{+} + n
\end{equation}
reaction. 
   In this reaction, both isospin $T$ = 0 and 1 states can be populated, while 
the reaction (\ref{eq:K-p-reaction}) is limited to forming $T$ = 1. 
   We describe the study of the neutron spectrum 
obtained from reaction (\ref{eq:K-n-reaction}) to  
provide further important information on the strange tribaryon.

\section{The experiment}

Let us briefly summarize the experiment while focusing on a 
neutron measurement. 
A more detailed description is given in Refs. \cite{Suz04,NIM01}.

\begin{figure}[hbt]
\begin{center}\includegraphics[width=0.8\columnwidth]{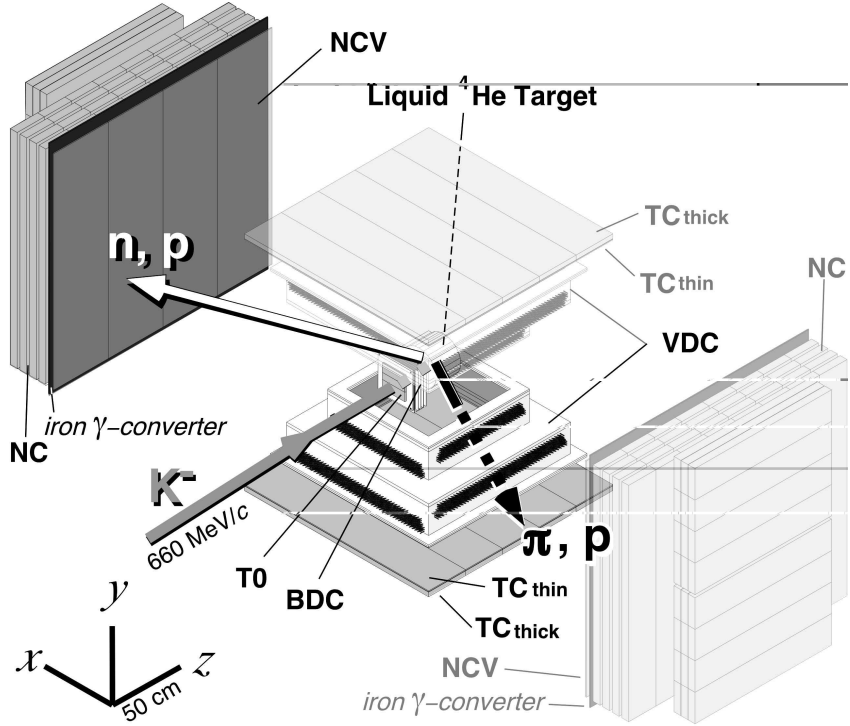}\end{center}
\caption{\label{fig:setup} 
	Schematic figure of the experimental setup. The notations are:  
	T0, beam timing counter; 
	BDC, beam-line drift chamber (16 layers);  
	VDC, vertex drift chamber (12 layers); 
	TC$_{\rm thin}$ and TC$_{\rm thick}$, 
	thin (0.6 cm) and thick (3 cm) trigger counter; 
	NCV, neutron-counter charged-particle veto; 
	NC, neutron-counter wall.
	To simplify, several counters are omitted.}
\end{figure}

   To obtain the neutron spectrum, we adopted 
the time-of-flight (TOF) method, 
as shown in Fig. \ref{fig:setup}.
   For this purpose, we need to define the kaon reaction point by  
the vertex between the trajectory of an incoming kaon and that 
of an outgoing charged particle detected by two layered trigger counters (TC), 
namely TC$_{\rm {thin}}$ and TC$_{\rm {thick}}$.
   The resolution of the reaction points/vertex was $\sim$ 5 $mm$ ($\sigma$),  
which is predominantly determined by kaon multiple scattering 
in the target helium.
   The kaon reaction at-rest in the target volume was selected (purity $>95\%$) 
by using the correlation between the pulse height 
of the final beam counter (T0) 
and the range in the helium target obtained by the vertex position. 
   The reaction (neutron TOF start) timing was calculated from the T0 timing
and the kaon range in the target, and the neutron was detected by 
two sets of plastic scintillation counter arrays (NC).
   An absolute time calibration of the TOF was performed 
by electromagnetic showers 
produced in iron-plate converters, placed between NC veto counters 
and NC, itself. 
   The overall TOF resolution was measured by this  
shower event, and was found to be $\Delta (1/\beta) = 0.04 (\sigma)$. 
   The absolute momentum was checked using mono-energetic neutrons from 
$\Sigma^+$ decay at rest ($\Sigma^+ \rightarrow n \pi^+$) by back-to-back 
coincidence with the threshold at 3 MeV $electron$ $equivalent$ (MeV$ee$) applied to NC.

   In a following analysis, we set a rather high energy threshold to the NC system 
(10 MeV$ee$) at the analysis stage, so as to reduce the accidental neutron background. 
   At this threshold, an NC counter of 5 cm thickness 
become sensitive at $\sim$ 150 MeV/$c$, and saturate at $\sim$ 340 MeV/$c$.
Beyond this, the efficiency is constant at $\sim$ 20\%.

\section{Analysis procedure}

\begin{figure}[hbt]
\begin{center}\includegraphics[width=0.9\columnwidth]{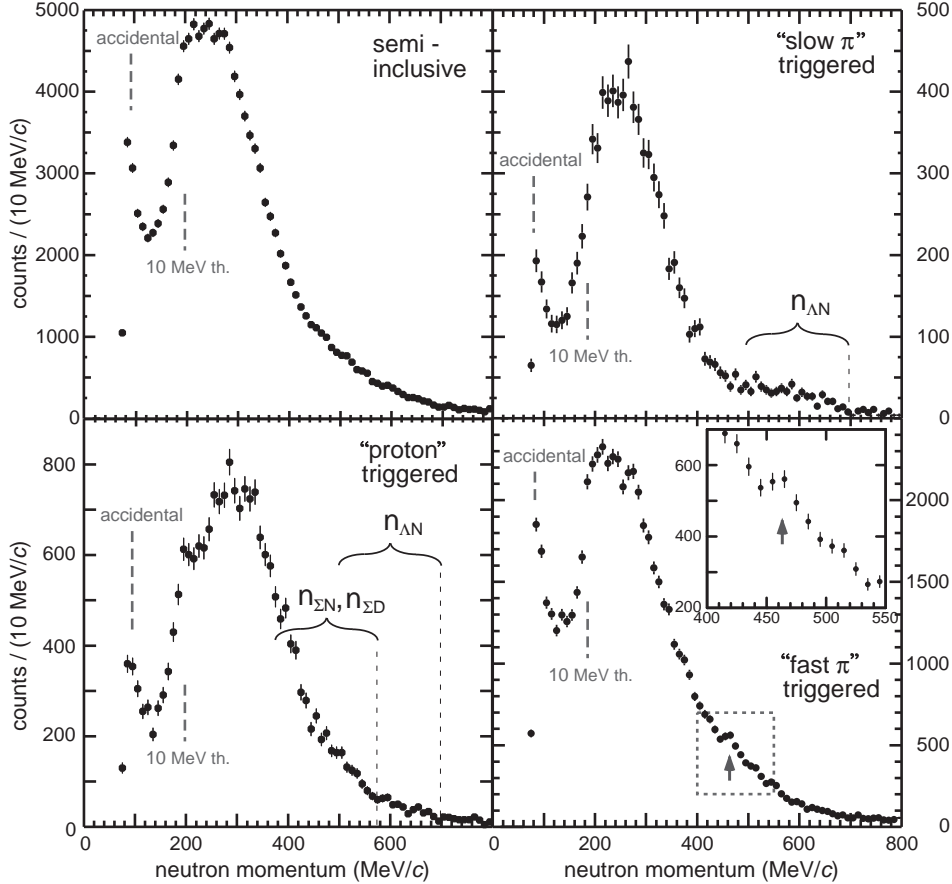}\end{center}
\caption{\label{fig:slow-fast-pi} 
	Neutron momentum spectra. 
	The upper-left panel shows the semi-inclusive spectrum without any event selection, 
	whereas the rest of the figures are subsets.
	The event selection of each subset is given in Fig. \ref{fig:dEdxcont}.
	A peak-like structure exists in a ``fast $\pi$'' triggered event set
	at around $\sim$ 470 MeV/$c$; a close-up view of the doted
	region is given as an inset.
}
\end{figure}

As shown in Fig. \ref{fig:slow-fast-pi} (upper left), 
the semi-inclusive neutron spectrum does not yield any distinct peak, 
in contrast to the proton spectrum \cite{Suz04}.

To study the neutron spectrum in more detail, 
we focused on pions that came from hyperon decay. 
focused on pions that came from
   If these pions come from hyperon decay, 
the mother hyperon could be qualitatively identified by the 
pion momentum. 
   The momentum of the pion from $\Lambda$ 
distributes dominantly in the lower momentum side (mostly below 100 MeV/$c$)
compared to that from a charged $\Sigma$ (centered at about 190 MeV/$c$) 
\cite{OUTA}. 

   For a further study, we proceeded to an analysis of the  
hyperon motion based on the vertex inconsistency. 
   If the hyperon moves a certain length,  
the two trajectories between the 
incoming kaon and the outgoing charged pion could be separated by the 
motion before the decay. 
   On the other hand, if the triggered pion come 
from the quasi-free hyperon production
 (QF : $K^-N \rightarrow \pi^\pm Y$), 
the vertex should coincide within the experimental resolution.

   If the main decay channel of ${\rm{S}}^{+}$ is similar to that 
of ${\rm{S}}^{0}(3115)$, then one can assume that the $\Sigma NN$ is 
the major decay channel. 
   Thus one expects  
\begin{equation}
\label{eq:S-decay}
	{\rm{S}}^{+} \rightarrow \Sigma^{\pm}NN
	\enskip \mathrm{and} \enskip
	\Sigma^{\pm} \rightarrow \pi^{\pm}N,
\end{equation}
which is an ideal model to apply the analysis procedure 
described above, because ${\rm{S}}^{+}$ is expected to be boosted 
at the formation stage.

\section{Neutron spectra classified by TC}

\begin{figure}[hbt]
\begin{center}\includegraphics[width=0.8\columnwidth]{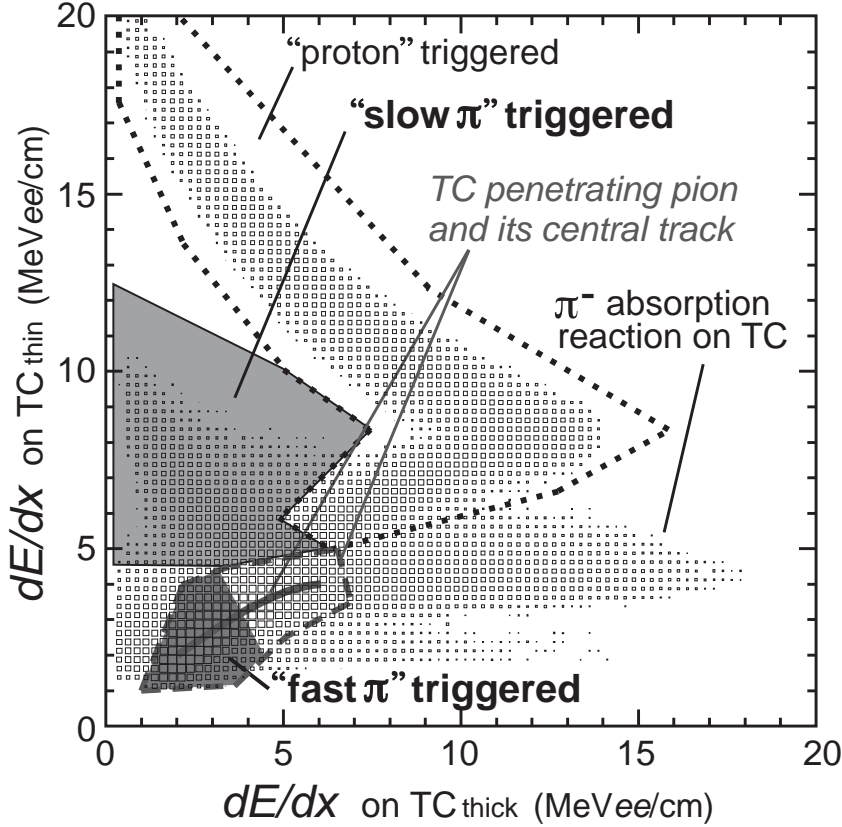}\end{center}
\caption{\label{fig:dEdxcont} 
	Scatter plot of dE/dx on TC$_{\rm {thick}}$ (horizontal-) 
	and TC$_{\rm {thin}}$ (vertical-axis) counters.
	Definition of ``proton'', ``slow $\pi$'' and ``fast $\pi$'' triggered events 
	are given in the figure.
}
\end{figure}

   Let us define the event window using the TC information.
   Fig.~\ref{fig:dEdxcont} shows the scatter plot of   
$dE/dx$ on two layers of the TC counters.
   To select high-momentum pions 
($p_\pi$ $\mathcomposite{\mathrel}{>}{\sim}$ 125 MeV/$c$), 
which mainly come from $\Sigma^\pm$ decay, we proceeded as follows.
   From the simulation, we know that the pions, which penetrate both layers
($p_\pi$ $\mathcomposite{\mathrel}{>}{\sim}$ 100 MeV/$c$), 
should distribute mostly inside the dashed region. 
   We evaluated the distribution center by using 
an average of the data within this region. 
   The central track obtained is shown in the figure as a solid line, 
and the evaluation of the momentum along the track 
led us to select the ``fast $\pi$'' triggered event window 
(hatched area in the dashed region). 
   This window is well apart from the ``proton'' window shown in the figure.

 The ``slow $\pi$'' triggered event window is defined so as to select 
pions that stop in the TC counters 
($p_\pi$ $\mathcomposite{\mathrel}{<}{\sim}$ 90 MeV/$c$).
Apart from the region shown in the figure, 
we also included a very slow pion which had already stopped in TC$_{\rm{thin}}$, 
because these events can be easily selected by the pulse height and TDC information.
   
   Fig.~\ref{fig:slow-fast-pi} shows the neutron momentum spectra for ``proton''
(lower left), ``slow $\pi$'' (upper right) and ``fast $\pi$'' (lower right)
triggered events tag. 
   There exists a common structure up to $\sim$ 400 MeV/$c$. 
   This low-momentum component mainly comes from the $\pi^-$ absorption reaction 
(occurring in or around the target),
$\pi^- NN \rightarrow NN$ \cite{piabs},  
and cascade reactions followed by QF-hyperon 
formation \cite{NIM01}.

   The ``slow $\pi$'' triggered neutron spectrum 
has a ``plateau'' structure  beyond $\sim$ 450 MeV/$c$. 
This comes from the kaon non-mesonic absorption and its chain,  
\begin{equation}
	K^-NN \rightarrow \mathnormal{\Lambda} n_{\Lambda \mathrm{N}}
	\quad \mathrm{and} \quad  
	\mathnormal{\Lambda} \rightarrow \pi^- p
\label{eq:KNN-LN},
\end{equation}
where $n_{\Lambda \mathrm{N}}$ is the neutron from the non-mesonic kaon 
absorption reaction. Note that pions from $\Sigma$ decays do not contribute
when we select low-momentum pions \cite{OUTA}.
   The spectrum shape of $n_{\Lambda \mathrm{N}}$ is dominantly determined 
by three-body phase space and the nuclear form factor, and is very broad, as shown 
in the figure.
   It has a maximum momentum of $\sim$ 700 MeV/$c$, 
and forms a ``flat-plateau'' below $\sim$ 600 MeV/$c$. 

   Similar to the ``slow $\pi$'' triggered events,
the neutron spectrum of ``fast  $\pi$'' triggered events is expected to have 
a high-momentum component due to
\begin{equation}
	K^-NN \rightarrow \mathnormal{\Sigma}^{\pm} n_{\Sigma \mathrm{N}} 
	(\mathrm{or} \enskip p)
	\quad \mathrm{and} \quad  
	\mathnormal{\Sigma}^{\pm} \rightarrow \pi^{\pm} n_{\Sigma \mathrm{D}} 
\label{eq:KNN-SN},
\end{equation}
where $n_{\Sigma \mathrm{N}}$ is the neutron at the primary reaction and 
$n_{\Sigma \mathrm{D}}$ is that from $\mathnormal{\Sigma}^{\pm}$ decay. 
   The $n_{\Sigma \mathrm{N}}$ should have a nature very similar to the
$n_{\Lambda \mathrm{N}}$ in reaction (\ref{eq:KNN-LN}),
but shifted by $\sim$ 100 MeV/$c$ to the lower side.
   Therefore, a ``plateau'' below $\sim$ 500 MeV/$c$ is expected.
   Actually, the ``proton'' triggered event is not selective to 
$n_{\Lambda \mathrm{N}}$, $n_{\Sigma \mathrm{N}}$ or $n_{\Sigma \mathrm{D}}$, 
as indicated in Fig.~\ref{fig:slow-fast-pi} (lower left).

   There exists a peak-like structure at $\sim$ 470 MeV/$c$
in the ``fast $\pi$'' triggered event (Fig.~\ref{fig:slow-fast-pi} (lower right)),
at the momentum where we expect a ``plateau'' due to 
$n_{\Sigma \mathrm{N}}$ and $n_{\Sigma \mathrm{D}}$. 
   Is this peak-like structure due to the formation of 
the intermediate state S$^+$? 
   As shown in the figure, 
the answer largely depends 
on what the background state is. 
and where the background level is around this region.

   If this peak-like structure indicates the formation of S$^+$, 
then the major decay will be $\Sigma NN$,
because the structure appears only in the ``fast $\pi$'' triggered spectrum. 
   To make the existence clear, we need to study the 
spectrum in more detail by analyzing the hyperon motion.

\section{Neutron spectra selected by the hyperon motion}

\begin{figure}[hbt]
\begin{center}\includegraphics[width=0.8\columnwidth]{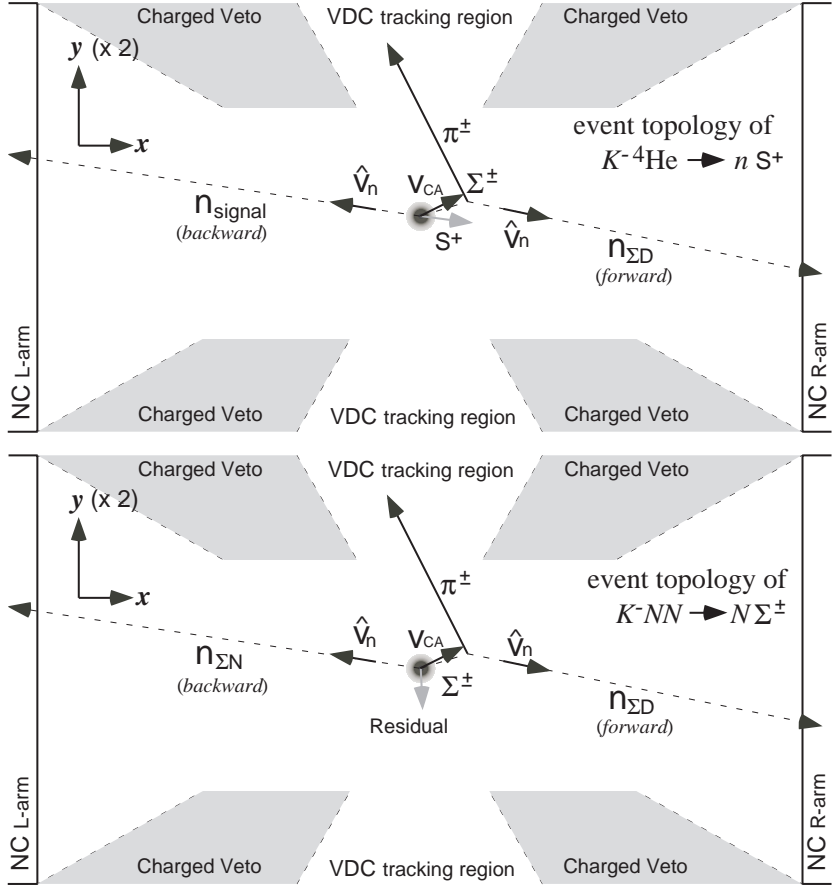}\end{center}
\caption{\label{fig:event3} 
	Example of the event topologies. The S$^+$ formation topology is shown at the top. 
	If the Q-value of S$^+$ formation is large, then the S$^+$ is boosted 
	to the opposite direction of the signal neutron, which results in the  
	${\bf v}_{\mathrm{CA}} \cdot {\bf \hat{v}}_n$ being distributed on the negative side.
	The background reaction (\ref{eq:KNN-SN}) is shown at the bottom.
	Although the two topologies are quite similar, the scalar product, 
	 ${\bf v}_{\mathrm{CA}} \cdot {\bf \hat{v}}_n$, 
	is negative (backward) if $n_{\Sigma \mathrm{N}}$ is detected
	by the neutron counter, while it is positive (forward) 
	for $n_{\Sigma D}$.
}
\end{figure}

    The most simple way to detect the hyperon motion is  
by a vertex inconsistency between the two tracks of 
an incoming kaon and an outgoing trigger particle 
(so called distance-of-closest-approach). 
   In the present case, however, the hyperon and the neutron 
motions are asymmetric in many reactions (see Fig.~\ref{fig:event3}).
   For a further analysis, let us introduce 
a scalar product, ${\bf v}_{\mathrm{CA}} \cdot {\bf \hat{v}}_n$, 
where ${\bf v}_{\mathrm{CA}}$ is the vector from the kaon trajectory to that of the 
charged pion at the minimum distance, and ${\bf \hat{v}}_n$ is the 
normalized vector of the detected neutron. 

\begin{figure}[hbt]
\begin{center}\includegraphics[width=0.8\columnwidth]{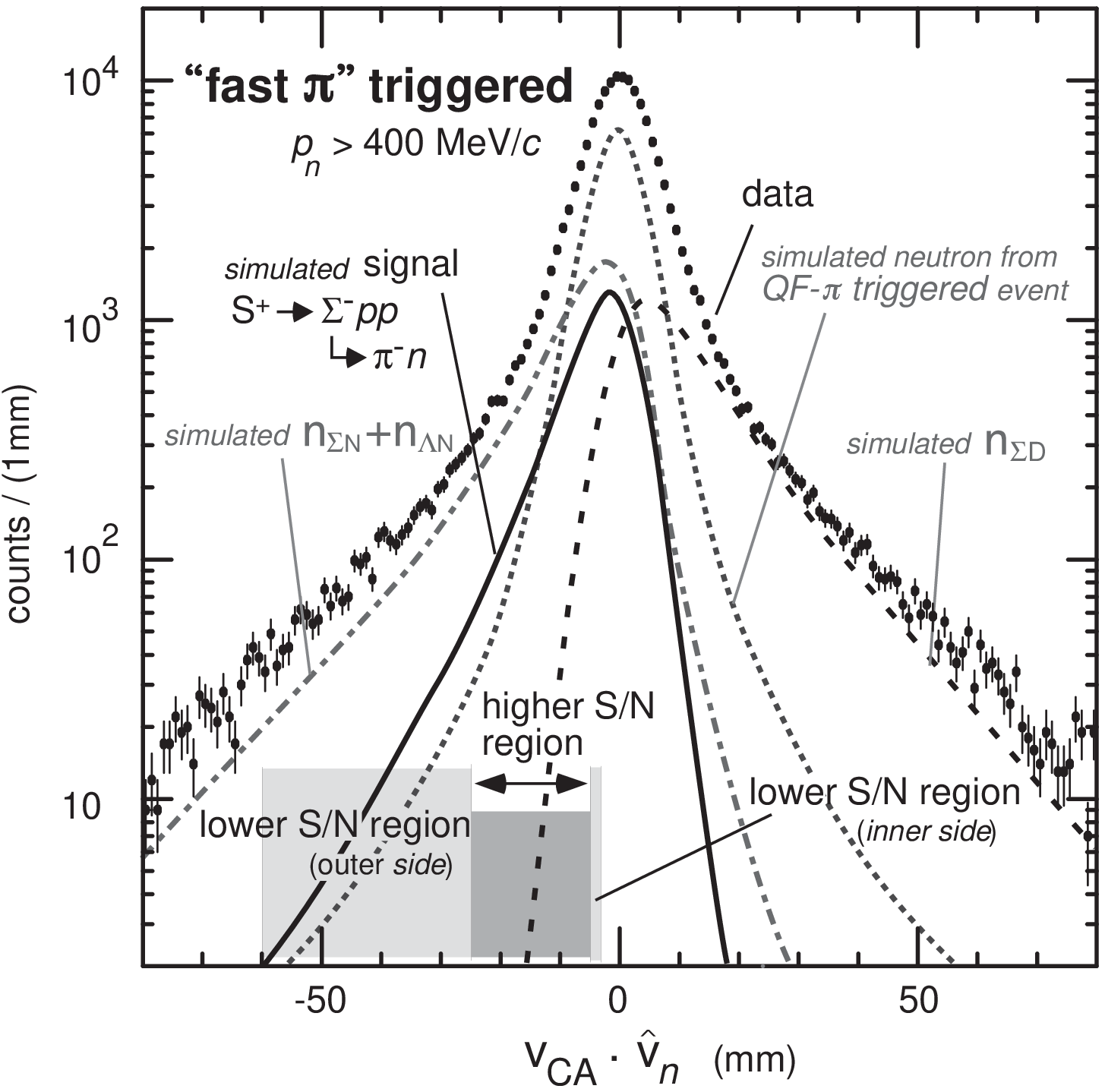}\end{center}
\caption{\label{fig:VcaVp} 
	${\bf v}_{\mathrm{CA}} \cdot {\bf \hat{v}}_n$ distribution.
	 The dots with error bars show the data of 
	the ``fast $\pi$'' triggered events 
	associated with a high-momentum neutron ($p_n >$ 400 MeV/$c$). 
	The decomposition into key components, given by the simulation, is also 
	shown.
 	To make the signal formation component visible, 
	the S$^+$ is simulated with a formation probability of 1\% per stopped $K^-0$,
	and decays purely into $\Sigma^{-}pp$, which is the most efficient branch to 
	detect in this ${\bf v}_{\mathrm{CA}} \cdot {\bf \hat{v}}_n$ selection.
}
\end{figure}

   Fig.~\ref{fig:VcaVp} shows the ${\bf v}_{\mathrm{CA}} \cdot {\bf \hat{v}}_n$
distribution of ``fast $\pi$'' triggered events having 
NC-detected neutron momenta $p_n >$ 400 MeV/$c$. 
   For a comparison, the decomposed spectra obtained by a simulation 
are shown.
   As shown in the simulation, $n_{\Sigma \mathrm{N}}+n_{\Lambda \mathrm{N}}$ and 
$n_{\Sigma \mathrm{D}}$ roughly distribute symmetrically on opposite sides.

   If the peak-like structure obtained in Fig.~\ref{fig:slow-fast-pi} (lower right) 
is due to the formation of an intermediate state, S$^+$, 
then it should be boosted opposite to the neutron with the same momentum, as shown in 
Fig.~\ref{fig:event3} (top).  
   It should then decay dominantly to
$\mathnormal{\Sigma} NN$ rather than $\mathnormal{\Lambda} NN$, 
as is the case for S$^0$, 
since we observe the structure only for a ``fast $\pi$'' triggered event.
   What is the centroid of the momentum of the decay $\mathnormal{\Sigma}^{\pm}$?
   We calculated the momentum, and found that it is about 450 MeV/$c$ 
in the backward direction, while it is about 80 MeV/$c$ in forward direction.
   Thus, the ${\bf v}_{\mathrm{CA}} \cdot {\bf \hat{v}}_n$ distribution 
for such events should be asymmetric and dominantly located on the negative side.
   This is the reason for the shape of the distribution of the simulated signal
given in Fig.~\ref{fig:VcaVp}.

   Unfortunately, there is no clear criterion about how to 
separate the signal from the background. 
   Simple ``forward'' and ``backward'' comparisons may not be valid, 
because the yield and the spectral shape can also be weakly asymmetric
between the ``forward'' ($n_{\Sigma \mathrm{D}}$) and 
``backward'' ($n_{\Sigma \mathrm{N}} + n_{\Lambda \mathrm{N}}$) directions.

   Therefore, we examined how to compare the event sets between 
higher S/N (HSN : indicated with arrow) and lower S/N (LSN : shallow hatched) regions, 
whose windows are defined as shown in the figure.
   HSN is defined to be where 
we expect a better signal-to-noise ratio than in the other regions. 
   The reason why we sandwiched HSN in between the two 
LSN regions (inner and outer side), is to obtain an interpolation of 
the shape of the spectrum. 
   It should be noted that both the spectrum shape and the statistics 
of the inner and outer LSN regions are different.
   If ${\bf v}_{\mathrm{CA}} \cdot {\bf \hat{v}}_n$ is close to zero, then 
the pion dominantly originates from the primary QF reaction, namely 
$K^-N \rightarrow \pi^\pm Y$. 
   If it is much apart from zero, 
the pion is dominantly from higher momentum hyperon decay, either from
reaction (\ref{eq:KNN-LN}) or (\ref{eq:KNN-SN}), 
which results in larger yields of the higher momentum component 
in the neutron spectrum. 

   It should also be noted that the LSN data are 
$absolutely$ $not$ free from the signal.
   In the case of the neutron spectra, the ratio of the signal yields 
between HSN and LSN
is expected to be about $\sim$ 2 : 1 from the simulation, 
if we assume that $\Sigma^\pm NN$ is the major decay branch of S$^+$.

\section{Discussion and Conclusion}

\begin{figure}[hbt]
\begin{center}\includegraphics[width=0.8\columnwidth]{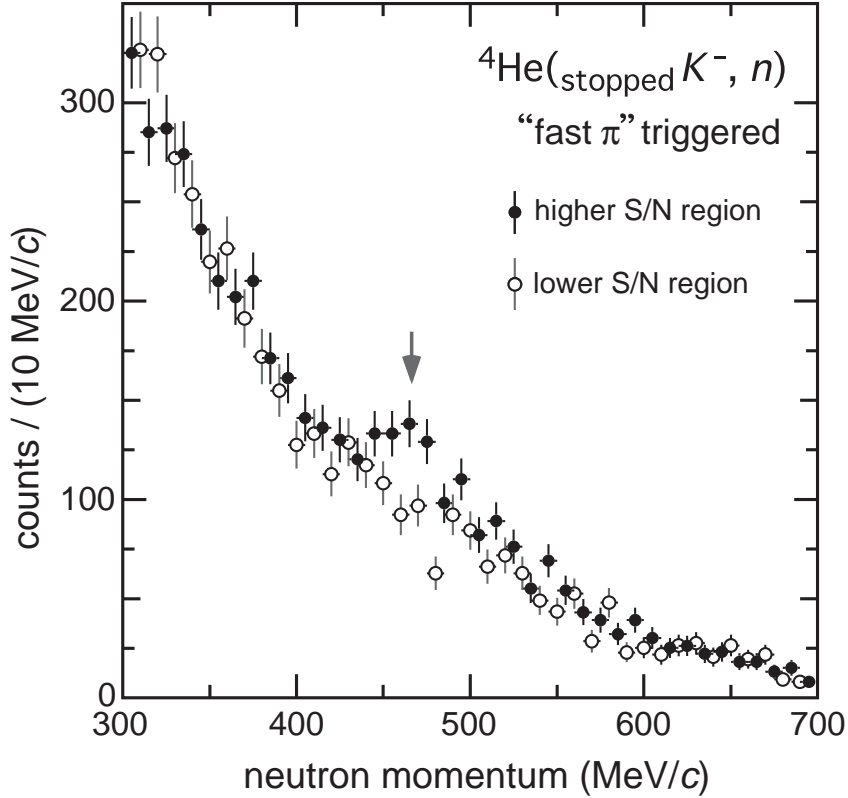}\end{center}
\caption{\label{fig:VcaVp_comp} 
	Comparison of the momentum spectra between HSN (closed circle) 
	and LSN (open circle). The LSN data were scaled so as to have the total yield 
	be the same as that of the HSN data, including the lower momentum region. 
	To avoid the error bars from overlapping, 
	the center of the bin is shifted by half of the bin size.
	The peak-like structure, observed in the ``fast $\pi$'' triggered neutron spectrum 
	(Fig.~\ref{fig:slow-fast-pi} (lower right)), appears as a clear excess between 
	HSN and LSN. 
}
\end{figure}

   Fig.~\ref{fig:VcaVp_comp} shows a comparison of the two spectra.
It exhibits a clear enhancement at around $\sim 470 $ MeV/$c$ 
in the spectrum of HSN, whereas it is smooth in LSN.
   Because the windows for HSN and LSN are 
rather arbitrarily defined, we also examined the stability of the enhancement 
by changing the cut region, and verified that it was nearly unchanged.
   The same spectra-shape comparison, whose windows were chosen to be 
symmetric to the present one 
in the ${\bf v}_{\mathrm{CA}} \cdot {\bf \hat{v}}_n$ positive side, 
was applied to check the validity of the comparison. 
   We found no clear excess, as was expected.

\begin{figure}
\begin{center}\includegraphics[width=0.8\columnwidth]{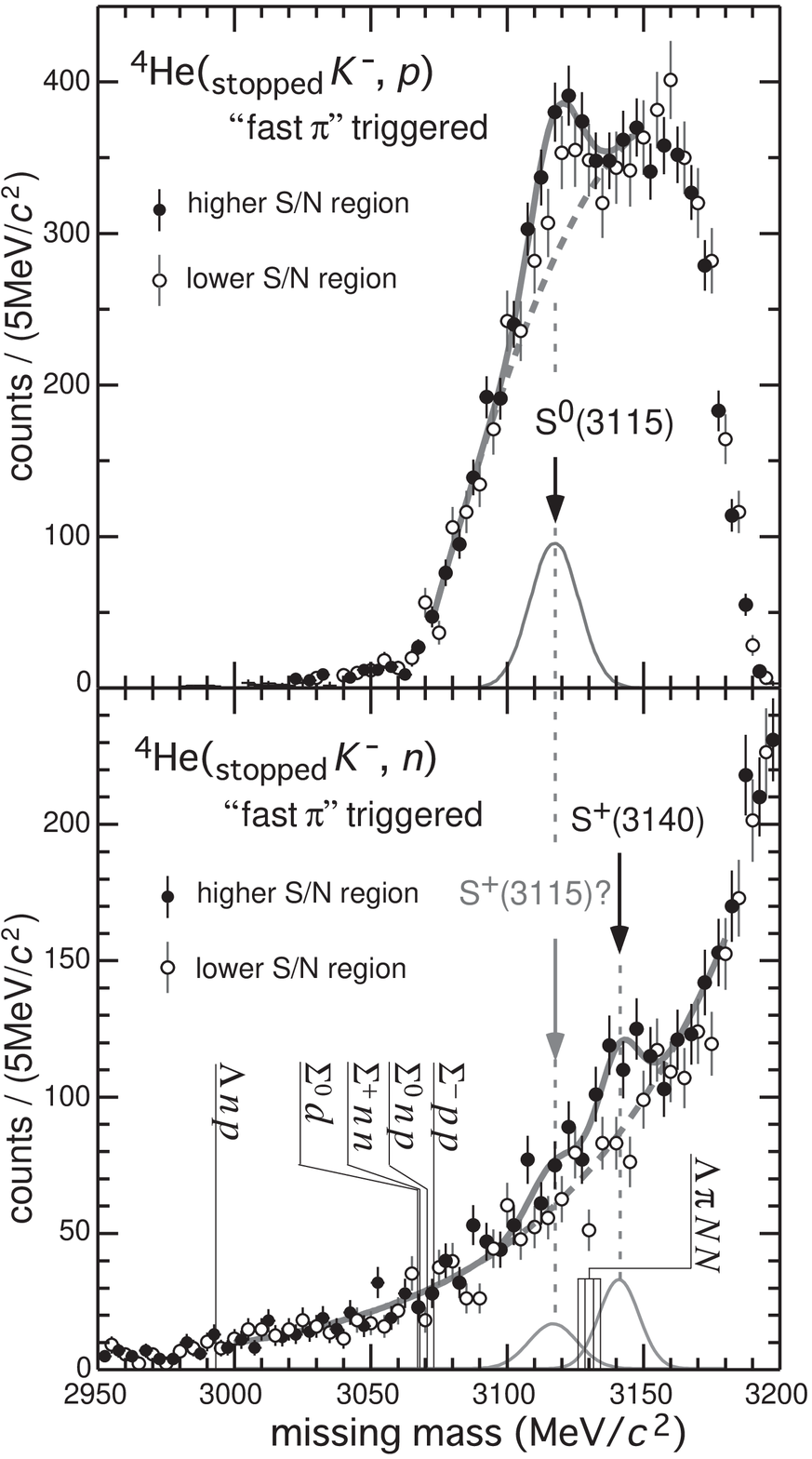}\end{center}
\caption{\label{fig:mass} 
   	Missing-mass spectra of the $^4$He$(_{\rm{stopped}}K^-, p)$ reaction (top)
	and the $^4$He$(_{\rm{stopped}}K^-, n)$ reaction (bottom). 
	Both proton and neutron HSN spectra were fitted 
	assuming a smooth background.
	 In the case of the HSN neutron spectrum (bottom), 
	we couldn't achieved a good fit 
	without assuming two Gaussian functions in the region of interest.
}
\end{figure}

   The missing-mass spectra both on HSN and LSN, 
obtained from the neutron data, are given in Fig.~\ref{fig:mass} (bottom). 
It shows that an enhancement exists at around the mass region 
from 3110 to 3160 MeV/c$^2$.
   As a third and final examination, 
we applied exactly the same procedure to obtain the mass spectra from 
proton data, as shown in Fig.~\ref{fig:mass} (top).
   We fitted the data of the HSN, 
assuming a smooth background for both the proton and neutron spectra. 
   The fit for the proton HSN spectrum (background shape of a 
third-order polynomial was used)
gave a consistent result with that obtained in a previous paper \cite{Suz04}. 

  The excess in the neutron spectrum seems to be rather broad compared to 
that in the proton spectrum. 
  It is natural to expect the isospin partner of ${\rm{S}}^{0}$(3115) 
in this energy region; hence, we fitted the HSN neutron spectrum with 
two Gaussian functions together with a smooth 
background (a single exponential), as shown in the figure, 
while fixing the mass and width of one 
Gaussian to be the same as that of ${\rm{S}}^{0}$(3115).
   In the fit, data of LSN were not utilized 
to constrain the background shape/yield. 
   Because the signal is expected to appear in both HSN and LSN 
at a ratio of 2 : 1, the background is expected to be 
located slightly below the LSN data.
   The result, however, is close to the data point of LSN, 
so that the fitted background slightly overestimated the 
yield in the region of interest.

   It is obvious that the statistical significance of 
the isospin partner of ${\rm{S}}^{0}$(3115)
is not enough to conclude its existence from the data itself.
   From the isospin rule, one can expect the 
formation ratio of the $T$ = 1 states between 
the neutron and the proton emission processes to be 1 : 2. 
   The lower neutron detection efficiency 
makes the ratio to be about 1 : 10. 
   If we assume that the decay branching ratio to the charged 
$\Sigma$ is the same from the two states, 
we expect the signal yield in the neutron spectrum to be 
about 1/10 of that in the proton spectrum,  
which is consistent even with the disappearance 
in the neutron spectrum.

   On the other hand, concerning the main signal component, 
the fit gave an energy of 3141 $\pm$ 3 MeV/$c^2$, width of 17 $\pm$ 5 MeV/$c^2$ 
and a yield of 120 $\pm$ 32 counts. 
   By folding the systematic error and the experimental resolution, 
we obtained a mass and width of 
$M_{\rm{S}^+} = $ 3141 $\pm$ 3 (stat.) $^{+4}_{-1}$ $(sys.)$ MeV/$c^2$
and $\mathnormal{\Gamma}_{\rm{S}^+} < 23$  MeV/$c^2$ (95\% $CL$), respectively.
   The significance of this excess, 
evaluated from the area of the Gaussian signal and its error, 
is 3.7 $\sigma$. The observed excess indicates the formation of 
another strange tribaryon (denoted as S$^+$(3140) hereafter), 
which lies about 25 MeV above S$^0$(3115). 
   Because of a lack of information about its 
decay branching ratio, we can not derive a reliable formation probability 
of S$^+$ from the fitted yield. 
   Based on simulations, the formation probability is below 
a few \% per stopped $K^-$; 
otherwise, the peak should already be observed in the semi-inclusive 
neutron spectrum.

   Since the mass of the signal is much different from S$^0$(3115), 
and also the yield is larger than the expected one, 
it is unlikely to be its isospin partner.
   This state could be assigned to the predicted $T=0$ state, 
although the observed mass is much lower than the 
theoretical one (3194 MeV). 

   In summary, we have found an experimental indication 
of another strange tribaryon S$^+$(3140) in the $^4$He(stopped-$K^-, n$) reaction. 
Its isospin is likely to be $T$ = 0, and the major 
decay mode is S$^+ \rightarrow \Sigma^{\pm} NN$. 
The present finding provides additional information about the strange tribaryon system. 
The most important finding is that both S$^0$(3115) and S$^+$(3140) 
have much smaller masses (namely, much larger $\bar{K}$ binding energies) 
than the theoretical values. 
   If these strange tribaryons, S$^0$(3115) and S$^+$(3140),
are indeed deeply bound kaonic states, 
then the huge total binding energies ($\sim$ 200 MeV) 
naturally imply the formation 
of a cold and dense nuclear system \cite{PRC02,DOTE}. 
   These findings provide important information about the 
${\overline{K}}N$ interaction, 
and may provide rich information to understand the 
dynamics of a dense system, such as a  
strange star \cite{quarkstar}. 
   In such a system, a restoration of chiral symmetry can be realized, 
and quarks may be deconfined \cite{chiral}, 
which are more or less unknown as of today.

\section*{Acknowledgments}

We are grateful to Y. Akaishi, R. Seki, D. Davis and G. Beer for fruitful discussions. 
We owe much to all the members of KEK-PS for their substantial cooperation. 
This work is supported by MEXT, KEK, Riken, KOSEF, NSF, DOE and KRF.


\begin{thebibliography}{00}

\bibitem{Suz04} T.~Suzuki et al., Phys. Lett. B 597 (2004) 263. 

\bibitem{PRC02} Y.~Akaishi and T.~Yamazaki, Phys. Rev. C 65 (2002) 044005.

\bibitem{NIM01} M.~Iwasaki et al., Nucl. Instrum. Meth. A 473 (2001) 286.

\bibitem{OUTA} H.~Outa et al., Prog. Theor. Phys. Suppl. 117 (1994) 177. 

\bibitem{piabs} Chun-Ming Leung et al., Letters to Il Nuovo Cimento, 11 (1969) 389.

\bibitem{DOTE} A.~Dot$\acute{\rm e}$, H.~Horiuchi, Y.~Akaishi and T.~Yamazaki, 
Phys.  Rev.  C 70  (2004) 044313.

\bibitem{quarkstar} B.~Freedman, L.D.~McLerran, Phys. Rev. D 17 (1978) 1109;  \\
G.E.~Brown et al., Nucl. Phys. A 567 (1994) 937.

\bibitem{chiral} T.~Hatsuda and T.~Kunihiro, Phys. Rev. Lett. 55 (1985) 158; \\
T.~Hatsuda and T.~Kunihiro, Phys. Lett. B 185 (1987) 304.

\end{thebibliography}
\end{document}